\pdfoutput=1

\documentclass[11pt,
a4paper,times,
numbered,print,index]{article}

\usepackage[parfill]{parskip}

\usepackage{setspace}
\usepackage{titlesec} 

\usepackage{color}
\usepackage{cancel}
\usepackage{soul}
\usepackage{dsfont}
\usepackage{graphicx}
\usepackage{placeins}
\usepackage{float}
\usepackage{amsmath,mathtools,leftidx}
\usepackage{amsfonts}
\usepackage{amssymb}
\usepackage{verbatim}
\usepackage{abstract}
\usepackage[hyperindex,breaklinks]{hyperref} %bookmarks=false
\hypersetup{
	colorlinks=false, %set true if you want colored links
	linktoc=all,     %set to all if you want both sections and subsections linked
	citebordercolor = {0 1 0},
	menubordercolor = {1 0 0},
}

\usepackage{subfigure}
\usepackage{authblk}
\DeclareGraphicsExtensions{.eps,.png,.jpg} %.png .pdf,
\usepackage{titlesec}

\titleformat{\chapter}[display]
{\bfseries\Large}
{\filleft\MakeUppercase{\chaptertitlename} \Huge\thechapter}
{1ex}
{\titlerule\vspace{1ex}\filleft}
[\vspace{1ex}\titlerule]

\titleformat{\section}
{\normalfont\sffamily\large\bfseries \Large}
{\thesection}{1em}{}

\titleformat{\subsection}
{\normalfont\sffamily\normalsize\bfseries \large}
{\thesubsection}{1em}{}

\titleformat{\subsubsection}
{\normalfont\sffamily\normalsize\bfseries}{\thesubsubsection}{1em}{}


\usepackage[backend=bibtex,style=phys,
articletitle=false,biblabel=brackets,
chaptertitle=false,pageranges=false,
natbib=true,arxiv=abs,eprint=true,maxnames=10,
backrefsetstyle=setonly,subentry,firstinits]{biblatex} %

\usepackage{geometry}

\newif\ifclasscode

\newenvironment{classcode}
{
	\global\classcodetrue
	\addvspace{30pt}%
	\small
	\noindent{{\bf{PACS:}} }\ignorespaces
}%{%
\newenvironment{keywords}
{
	\small
	\noindent{{\bf{Keywords:}} }\ignorespaces
}

\allowdisplaybreaks

\makeatletter
\g@addto@macro\normalsize{%
	\setlength\abovedisplayskip{15pt}
	\setlength\belowdisplayskip{15pt}
	\setlength\abovedisplayshortskip{15pt}
	\setlength\belowdisplayshortskip{15pt}
}
\makeatother
\begin{document}

%\title{\textit{Gravity in the Quantum Lab}}

\title{\sffamily\huge \textbf{Gravity in the Quantum Lab} \vskip13pt}

\author[$\dagger$]{Richard Howl\thanks{Corresponding author, email: richard.howl@univie.ac.at}}
\author[$\ddagger$]{Lucia Hackerm\"uller}
\author[$\mathsection$]{David Edward Bruschi}
\author[$\dagger$,$\ddagger$]{Ivette Fuentes}
\affil[$\dagger$]{\small \textit{University of Vienna, Boltzmanngasse 5, 1090 Wien, Austria}}
\affil[$\ddagger$]{\small \textit{University of Nottingham, University Park, Nottingham NG7 2RD, United Kingdom}}
\affil[$\mathsection$]{\small \textit{York Centre for Quantum Technologies, Department of Physics, University of York, YO10 5DD Heslington, UK}}

\date{}

\maketitle

\setstretch{1.2}

\vskip10pt

%------------------------------------------------------------------------------------------------------------------------------------------------------------------%
\begin{abstract}
%------------------------------------------------------------------------------------------------------------------------------------------------------------------%
\vskip5pt

At the beginning of the previous century, Newtonian mechanics fell victim to two new revolutionary theories, Quantum Mechanics (QM) and General Relativity (GR).  Both theories have transformed our view of physical phenomena, with QM accurately predicting the results of experiments taking place at small length scales, and GR correctly describing observations at larger length scales. However, despite the impressive predictive power of each theory in their respective regimes,  their unification still remains unresolved.  Theories and proposals for their unification exist but we are lacking experimental guidance towards the true unifying theory.  Probing GR at small length scales where quantum effects become relevant is particularly problematic but recently there has been a growing interest in probing the opposite regime, QM at large scales where relativistic effects are important.  This is principally due to the fact that experimental techniques in quantum physics have developed rapidly in recent years with the promise of quantum technologies. Here we review recent advances in experimental and theoretical work on quantum experiments that will be able to probe relativistic effects of gravity on quantum properties, playing particular attention to the role of Quantum Field Theory in Curved Spacetime (QFTCS) in describing these experiments.  Interestingly, theoretical work using QFTCS has illustrated that these quantum experiments could be used to enhance measurements of gravitational effects, such as Gravitational Waves (GWs). Furthermore, verification of such enhancements, as well as other QFTCS predictions  in quantum experiments, would  provide the first direct validation of this limiting case of quantum gravity, several decades after it was initially proposed.

%------------------------------------------------------------------------------------------------------------------------------------------------------------------%
\end{abstract}
%------------------------------------------------------------------------------------------------------------------------------------------------------------------%

\setstretch{1}

\begin{classcode}
04.80.-y, 03.67.-a, 04.62.+v, 03.70.+k
\end{classcode}


\thispagestyle{empty}

\newpage

\clearpage
\setcounter{page}{1}

\tableofcontents

%------------------------------------------------------------------------------------------------------------------------------------------------------------------%
\section{Introduction}
%------------------------------------------------------------------------------------------------------------------------------------------------------------------%
We all have an intuitive notion of what time and space are. In our experience, time flows and we see objects occupying a place in space. The physics of these scales of space and time that we experience are described by Newton's laws. The fundamental workings of amazing technologies, such as computers and mobile phones, are all based on these classical laws.\footnote{For example, a computer relies on classical information theory and stores information in binary bits so that, even though an understanding beyond Newtonian physics is required to manipulate certain components of the technology that are used to carry out the classical information tasks, the computer is fundamentally based on classical concepts.}   Interestingly, at different space and time scales, reality seems very different and it is not in harmony with our experience. On the one hand, physical laws are governed  by GR at large spatial scales, while on the other hand, QM rules the microscopic world. 

These theories have been extensively demonstrated by experiments. The state of systems, such as atoms and electrons, is governed by 
the wave-function, according to which quantum systems can occupy different positions at the same time.\footnote{This is just one of many possible interpretations of the physical meaning of the wave-function.} Another property of quantum systems is that multipartite systems can be in entangled states in which measurements in one system immediately affect the state of the other systems. A question of great interest in the field of QM is at which point physical systems lose their quantum properties and start behaving classically. Typically, systems consisting of a large amount of particles behave classically. However, in the case that such particles are at very low temperatures, they can exhibit quantum behaviour. This is the case of Bose-Einstein condensates (BECs): they consist of around $10^6$ atoms cooled down to the nano-Kelvin regime. Excitingly, we are now entering a new era in which we are starting to develop technologies based on the physics of the microscopic world. Quantum technologies, including quantum communication and quantum metrology, have become popular because they enable us to achieve tasks that are not possible in a classical context. The last decades have seen accelerated development of these technologies, including the control of systems for which Haroche and Wineland were awarded the 2012 Nobel Prize in Physics. 

Let us now discuss physics at large length scales. In GR, space and time form part of a more general structure that is four-dimensional and is called spacetime. Contrary to our experience, where space and time are absolute notions, spacetime is relative.  Einstein taught us that the notion of spacetime could help us understand gravity as the curvature of spacetime produced by the presence of energy. GR also predicts the existence of Gravitational Waves (GWs). These are spacetime distortions produced by certain accelerated energy distributions, such as supernova explosions or moving black holes. After almost one hundred years since they were first predicted, GWs were finally observed for the first time this year by the Laser Interferometer Gravitational-Wave Observatory (LIGO) in a milestone moment in physics \cite{PhysRevLett.116.061102}. However, we are still lacking experiments that help us understand GR at small lengths or large energies where quantum effects become relevant.  Our understanding of the interface of these two fundamental theories is  incomplete and the inability to unify them remains one of the biggest unsolved problems in physics.  

Understanding GR at small length scales where quantum effects become relevant is a highly non-trivial endeavour that suffers from a lack of experimental guidance. An alternative approach is to study quantum effects at large scales where experiments promise to be within reach in the near future. For example, one could study whether entanglement is affected by general relativistic effects such as time dilation in a gravitational field. Indeed, quantum experimental techniques have developed quickly in the last decades and are now reaching relativistic regimes.  This is illustrated, for example, by quantum teleportation being successfully implemented across 143\,km in the Canary islands, and long-range quantum experiments are under way to distribute entanglement across thousands of kilometres between Earth and satellite-based stations. At these regimes relativistic effects kick in. For example, the Global Positioning System (GPS), a system of satellites used for time dissemination and navigation, requires relativistic corrections to determine time and positions accurately. However, we know very little about how gravity and motion affect quantum properties, such as quantum coherence and entanglement, that are used in quantum technologies. Relativistic regimes are being reached not only by space-based experiments but also through boundary conditions moving at relativistic speeds in superconducting circuits. These systems have been used to demonstrate relativistic quantum field effects in flat spacetime such as the Dynamical Casimir Effect (DCE). Furthermore, extremely high precision in quantum experiments has also enabled experimentalists to reach relativistic regimes by, for example, being able to prove time dilation at distances as small as a few cm where previous experiments assumed spacetime to be flat. By being able to measure space and time more precisely using quantum systems, we will soon be able to probe new regimes where new physics might emerge. 

In order to predict the results of quantum experiments at relativistic regimes we require a consistent theoretical framework. Fortunately, we do have a theory that enables the study of the overlap of these theories at low energies.  This is QFTCS and describes the behaviour of quantum fields in a classical curved spacetime. Predictions of the theory include Hawking radiation and particle creation by an expanding universe.  There are also analogous effects such as the Unruh and DCE generated from QFT in flat spacetime with accelerating reference frames.  Although the latter flat spacetime effect was demonstrated in 2011 using superconducting circuits \cite{Wilson2011}, QFTCS still awaits experimental demonstration despite most paradigmatic predictions of the theory being over three-decades old. Due to obvious difficulties in demonstrating Hawking radiation and particle creation by an expanding Universe, several groups have been working in analogue demonstrations of these effects. In analogue gravity, researchers find ways of reproducing the basic features of the theory in systems such as BECs, water waves, non-linear optical waveguides and superconducting circuits. These experiments cannot strictly falsify QFTCS since they are based on analogue effects, instead they are probing the mathematics of the theory and can only find potential clues to the breakdown of the physical theory. However, new proposals using QFTCS have identified \textit{real} spacetime effects on quantum fields that could, in principle, be within experimental reach.  As well enabling a number of experiments in the overlap of quantum theory and GR, this will ultimately also lead to the development of technologies that are compatible with spacetime while exploiting both quantum and relativistic effects. These technologies have so far been suggested to be, in principle, implementable in BECs, superconducting circuits and free space quantum communication experiments.  For example, recent results have suggested that the phononic field of a BEC is particularly well-suited to measure spacetime effects. Although BECs can be as small as one micrometer, they can be cooled down to a few nano-Kelvin where they not only exhibit fascinating quantum behaviour but are also, surprisingly sensitivity to very weak spacetime changes. This new observation enables the possibility of developing a microscopic device capable of probing spacetime effects directly.  Technology of this kind would open a new window promising to provide deeper insights about our Universe through gravity.   A principle aim is to use these effects on a BEC to develop a GW detector capable of detecting spacetime distortions produced by merging neutron stars and black holes as well as rotating neutron stars and pulsars.  This would offer a much cheaper and practical GW detector than the current state of the art.  Multiple GW detectors could then be used by various experimental groups, allowing for improved validation of the signal through simultaneous measurements, as well as superior positioning of the source of the signal.

The main aim of these relativistic quantum technologies is to develop new designs that not only exploit quantum properties but also take advantage of, or incorporate, spacetime effects. These new devices are being actively investigated in the research field of relativistic quantum information and metrology.  In this field, research groups try to develop quantum technologies that are compatible with relativity, and some ideas for using relativistic effects to improve quantum technologies, such as quantum measurement technologies \cite{Ahmadi:Bruschi:2014:v2}, have surfaced.  The application of relativity in the development of new technologies has remained less explored than its quantum counterpart, but these new proposals illustrate the potential benefits that can be gained from utilising relativity. 

The rest of this work is organized as follows.  In Section \ref{sec:QMExpsRelativiticRegimes} we review  quantum experiments that are approaching regimes in which relativistic effects due to gravity will  become relevant.  In general, these experiments are designed and described using non-relativistic QM.  However, in Section \ref{sec:QFTCS} we discuss the motivation for moving to a QFTCS framework to describe such experiments. We also discuss other motivations for working with QFTCS, such has how this creates new possibilities in technology and how it could shed new light on answering fundamental questions relevant to quantum gravity. The theoretical work that has illustrated the need for using QFTCS is then reviewed in more detail in Sections \ref{sec:corrections}-\ref{sec:RQTechnology}. Specifically, in Section \ref{sec:corrections} we review recent works in QFTCS that have shown how the performance of certain quantum information protocols can be affected by relativistic effects, requiring corrections to be made to the predicted outcomes of the experiments.  The measurement of such effects would also represent an observation of QFTCS and, in the same section, we additionally review proposals that have been purposely designed to illustrate these effects using current technology rather than relying on future experiments designed to advance quantum technology.  Section \ref{sec:RQTechnology} then reviews recent proposals based on QFTCS that demonstrate how a new type of technology could be developed that exploit the relativistic effects, in contrast to current quantum technologies that would see the effects as a hindrance to their design. As well as providing greater performance than their non-relativistic counterparts, the implementation of such devices would also provide an observation of QFTCS effects and offer further insight into the interplay of quantum physics and GR.  Finally, in Section \ref{sec:Summary}, we conclude this review and discuss future prospects. While this is not an exhaustive review of theoretical and experimental works related to relativistic effects on quantum properties, we hope that it provides an overview of the current status of this exciting field, and that it may inspire new research.

%------------------------------------------------------------------------------------------------------------------------------------------------------------------%
\section{Quantum Experiments Approaching Relativistic Gravity Regimes} \label{sec:QMExpsRelativiticRegimes}
%------------------------------------------------------------------------------------------------------------------------------------------------------------------%

In this section we review quantum experiments that are approaching regimes where quantum properties will be modified by relativistic effects of gravity.  Generically, these quantum experiments are approaching such regimes because they involve large distances and implement high precision techniques.  Studying relativistic effects however is not, in general, the principal motivation behind the development of most of these quantum experiments.  Instead the experiments are primarily designed to advance quantum technology and its applications, such as the quantum internet, navigation, geophysics and GW astronomy.  Otherwise they have been designed to advance our understanding of fundamental physics, such as a modifications to quantum theory at large scales, or the quantization of gravity. 

In the following subsections we review the following types of quantum experiments that are approaching relativistic gravity regimes: large-scale experiments that are designed to advance quantum communications; experiments using ultra-precise quantum clocks; quantum technologies that are being used in gravitational metrology; quantum experiments that can detect GWs; quantum experiments that are specifically searching for deviations from GR; experimental proposals designed to search for quantum gravity using massive quantum systems; and quantum experiments that have exhibited the DCE, which is an analogue of effects due to curved spacetime.  

%All of these types of quantum experiments are reviewed in  Sections \ref{sec:LongRange}-\ref{sec:TestingGR} and Section \ref{sec:DCE}.  

%------------------------------------------------------------------------------------------------------------------------------------------------------------------%
\subsection{Long-range Quantum Communication Experiments} \label{sec:LongRange}
%------------------------------------------------------------------------------------------------------------------------------------------------------------------%
A primary goal of quantum technology is to develop a global quantum communications network and, ultimately, a quantum internet, which would provide an exponential speed-up in distributed computation and be impenetrable to hackers.  To achieve this goal it is likely that quantum systems will need to operate over large distances such as that of the Earth to satellites.  These are length scales where the relativistic nature of the Earth's gravitational field already needs to be taken into account in classical technologies such as GPS signalling.  Therefore, one would naturally expect that relativistic effects on quantum properties of long-range quantum communications experiments will also need to be considered.  Indeed, in Section \ref{sec:corrections}, we review theoretical works which demonstrate that these effects could  be crucial to the workings of such experimental setups.

An important step in the realisation of a global quantum communications network was made in 2011 when the group led by Anton Zeilinger distributed entanglement across a 144\,km free-space link between Canary islands La Palma and Tenerife \cite{Ma2012,Herbst17112015}. This set the record for the distance achieved for quantum entanglement and teleportation, and provided the benchmark for an efficient quantum repeater, which could be placed at the heart of a global quantum-communication network.  In collaboration with the Chinese Academy of Sciences (CAS), the group is now planning to demonstrate satellite-based quantum teleportation, which is anticipated to take place later this year. 

Quantum transmission is also being investigated between the Earth and a satellite. In particular, in 2015 a single photon exchange was performed from a medium Earth orbit satellite  to the ground station at the Matera Laser Ranging Observatory \cite{PhysRevLett.115.040502,Dequal:Vallone:2016}.  In this setup, a single photon source in orbit was simulated by emitting a train of laser pulses  with average photon number per pulse close to one, which were then reflected back at the single photon level from the satellite. The repetition rate of the laser pulses was 100\,MHz and the satellite was at more than 7000 km of slant distance.  As well as providing a further step towards a global communication network, this experiment also opens up the potential for fundamental tests of Quantum Mechanics with moving frames, and  \cite{Dequal:Vallone:2016}.  Furthermore, with an upgrade of the detector on ground, the group claim that it would be possible to achieve quantum communication for up to 23000 km \cite{Dequal:Vallone:2016}. In 2015 the group also demonstrated interference at the single photon level along satellite-ground channels in a step towards observing the gravitational phase shift between a superposition of two photon wave packets in a large distance quantum optics experiment in space \cite{PhysRevLett.116.253601}.

%%%%%%%%%%%% MORE DETAIL %%%%%%%%%%%%%
Many other groups are working on quantum communication between Earth and satellite links.  For example, the Institute for Quantum Optics and Quantum Information (IQOQI) in Austria is planning on performing quantum optics experiments, such as a Bell-type experiment with an entangled pair of photons, using an optical ground station (OGS) and the International Space Station (ISS) \cite{Scheidl2013};  the quantum information processing group (QIV) at the Max Planck Institute for the Science of Light (MPL) is investigating quantum communication experiments between the Teide Observatory on Tenerife and the Alphasat I-XL satellite at around a separation of 36000\,km \cite{Elser2015}; the Centre for Quantum Technologies (CQT) group at the National University of Singapore, together with the Satellite Research Centre at Singapore's Nanyang Technological University, is investigating a Small Photon-Entangling Quantum System (SPEQS) that will be set into orbit on a type of small satellite known as a CubeSat and be used as a testbed for technology for future quantum communication networks \cite{Chandrasekara2015,Bedington2015,PhysRevApplied.5.054022}; the National Institute of Information and Communications Technology (NICIT) in Japan is studying a microsatellite mission called SOCRATES with a goal to experiment with Quantum Key Distribution (QKD) techniques using a Small Optical Transponder (SOTA) on board a small satellite and an optical ground station located at NICT \cite{Takenaka2011}; the Institute for Quantum Computing (IQC) at the University of Waterloo, in collaboration with industry partners, %the Space Flight Laboratory (SFL) at the University of Toronto Institute for Aerospace Studies (UTIAS), 
have proposed a microsatellite mission Quantum Encryption and Science Satellite (QEYSSat)  to demonstrate the generation of encryption keys through the creation of quantum links between ground and space, and to conduct fundamental science investigations of long-distance quantum entanglement \cite{Jennewein2014}; and the Chinese Academy of Sciences (CAS) is investing heavily in projects on long-distance satellite and ground quantum communications \cite{PanJianwei547}, for example, see \cite{Yin:13}.

%%%%%%%%%%%% LESS DETAIL %%%%%%%%%%%%%
%Many other groups are working on quantum communication between Earth and satellite links, including, for example, the Institute for Quantum Optics and Quantum Information (IQOQI) in Austria \cite{Scheidl2013}; the quantum information processing group (QIV) at the Max Planck Institute for the Science of Light (MPL) \cite{Elser2015}; the Centre for Quantum Technologies (CQT) group at the National University of Singapore \cite{Chandrasekara2015,Bedington2015,PhysRevApplied.5.054022}; the National Institute of Information and Communications Technology (NICIT) in Japan \cite{Takenaka2011}; the Institute for Quantum Computing (IQC) at the University of Waterloo \cite{Jennewein2014}; and the Chinese Academy of Sciences (CAS), which is investing heavily in long-distance satellite and ground quantum communications \cite{PanJianwei547}, for example, see \cite{Yin:13}.

%------------------------------------------------------------------------------------------------------------------------------------------------------------------%
\subsection{Quantum Clocks} \label{sec:QMClocks}
%------------------------------------------------------------------------------------------------------------------------------------------------------------------%

The previous section reviewed quantum experiments that operate over large enough distances such that effects from the gravitational field of the Earth start to play a role.  However, with increasing precision of an experiment, the length scales over which one must consider general relativistic effects will significantly reduce.  This was dramatically demonstrated in the optical atomic clock experiment performed by the group of David Wineland %(Boulder) 
where  the time dilation due to the Earth's gravitational field was observed over a change of height of just 33\,cm \cite{Chou1630}.  This was achieved by comparing the  frequencies of two  $\mathrm{Al}^{+}$ clocks at two sample height differences. %The clocks exhibit a fractional frequency change of (4.1 T 1.6) × 10−17, which was measurable in this experiment with a height change of just 33\,cm since frequency variations below 10−16  in the clocks due to relativistic effects can be observed.  
However, although these type of aluminium-based clocks demonstrate remarkable accuracy, they are currently not the most accurate time keepers. Instead optical lattice clocks currently hold this mantle with a  $^{87}\mathrm{Sr}$ optical lattice clock showing a total uncertainty of $2.1 \times 10^{-18}$ in fractional frequency units \cite{Nicholson2015}.  This would correspond to a measurable gravitational time dilation for  a height change of just 2\,cm on the Earth \cite{Nicholson2015}.  

In addition to time dilation due to the Earth's gravitational field, the aluminium-based atomic clock was also used in a separate experiment to simulate the twin paradox and to observe time dilation for relative speeds of less than 10\,$\mathrm{ms}^{-1}$.   For this experiment the two ion clocks were placed in different labs, spaced 75\,m apart and connected via a stabilized optical fibre. One atom, the one corresponding to the travelling twin, was then set into an oscillating motion around its equilibrium position. Time dilation for this motion leads to a fractional frequency shift for the moving clock, which was found to be in agreement with that expected from relativity.

It is remarkable that general relativistic effects can be seen at such small length scales where one would naturally expect spacetime to appear completely flat, illustrating the power and precision of current technology.  Furthermore, these length scales are expected to reduce in the near-term with future major developments such as the utilisation of quantum correlations and implementations in space.  In the former case, entanglement could be used to enable quantum clocks operating close to the Heisenberg limit \cite{PhysRevLett.112.190403}, the ultimate fundamental bound imposed by QM, whereas, in the latter case, space-based operations would significantly reduce noise, allowing for new applications in fundamental physics, geophysics, astronomy and navigation.

The results from these experiments can so far be explained by modelling the clocks as  classical point-like, time-keeping systems in a relativistic gravitational field.  However, in Section \ref{sec:corrections} it is argued that, as the precision continues to improve, and as quantum correlations such as entanglement begin to be utilized, relativistic effects will start to influence the quantum properties of quantum clocks such that the results of experiments cannot be solely explained using this semiclassical approach.  As well as modifying the precision of quantum clocks, this would also enable  such clocks to be used as probes to study relativistic effects on quantum properties. In principle, the most appropriate and accurate theory for describing these quantum clock experiments that operate in relativistic regimes is QFTCS rather than non-relativistic QM since time is absolute in the Schr\"{o}dinger equation, which is a notion that is incompatible with the concept of time dilation.

%------------------------------------------------------------------------------------------------------------------------------------------------------------------%
\subsection{Gravitational Metrology Using Quantum Systems} \label{sec:GMetrology}
%------------------------------------------------------------------------------------------------------------------------------------------------------------------%

In the last few decades quantum technology, specifically atom-interferometry, has transformed the field of gravitational metrology. In particular, atom interferometers have found major applications in precise measurements of gravitational acceleration \cite{Kasevich1992,Peters1999,A.PetersK.Y.Chung2001}, gravity gradients \cite{PhysRevLett.81.971,PhysRevA.65.033608} and as gyroscopes \cite{PhysRevLett.78.2046,Gustavson2000}. These experiments all utilize the wave nature of atoms by sending them through an interferometer and  measuring the difference in phase induced by different gravitational potentials, an effect that was first demonstrated with neutrons in the Colella-Overhauser-Werner (COW) experiment in 1975 \cite{PhysRevLett.34.1472}.    For example, in a  trend-setting experiment by Guglielmo Tino (LENS) \cite{PhysRevLett.114.013001}, the change in the gravity gradient (known as the `curvature') of the gravitational field induced by test masses arranged around the experimental setup was measured using three atomic interferometers in parallel. The experiment used cold $^{87}$Rb atoms and relied on the combination of three vertical Mach-Zehnder interferometers implemented via Raman beams. Three clouds of cold atoms with a temperature of $\sim 4\,\mathrm{\mu K}$ were prepared in a superposition of the $\mathrm{F}=1$ and $\mathrm{F}=2$ states using counter-propagating Raman beams. The three clouds were then launched to different heights such that each trajectory corresponded to an interferometer with a different length. 
%A 3D fit to the resulting phase-curves rejects common noise and is used to reveal the curvature of the gravitational field.  
They obtained a gravity curvature of $\approx 1.4\times10^{-5}$ s$^{-2}$m$^{-1}$ for the arranged test masses, which are well-characterized tungsten weights with a total mass of about 519\,kg, and this agreed with a Monte-Carlo simulation to an accuracy on the per mill level.

Another example is the large scale Matter-Wave laser Interferometer Gravitation Antenna (MIGA), which consists of several spatially separated atom interferometers horizontally aligned and interrogated by a common laser field inside a 200-meter-long optical cavity.  This detector is currently being built in a low-noise underground laboratory in France \cite{refId0,R.Geiger2015} and is designed to provide measurements of sub Hertz variations of the gravitational strain tensor. Such light-pulse atom interferometers have demonstrated great potential as highly accurate quantum sensors, and recently a representation-free description of such devices was achieved that should allow future studies to consider arbitrary trajectories and varying orientations of the interferometer setup in non-inertial frames of reference \cite{Kleinert20151}.

Note that these atom interferometer experiments are testing for the effects of gravity on quantum properties of the system, namely quantum coherence%as superposition that have no classical-counterpart  
.  To date only effects that are fully accountable using Newtonian gravity have been observed.  %So far the experimental realizations of atom interferometers that measure gravity have been shown to be competitive with existing `classical' devices. 
However, major advances in the precision of these devices are expected in the future with, for example, the exploitation of highly entangled states.  There are also plans to use atom interferometers in space since certain ground-based systematic effects will be suppressed, allowing for improved sensitivity of the devices by orders of magnitude \cite{Tino2007159,Sorrentino2011}.  With these advances, effects from GR could become observable and would then need to be accounted for in their operation.  In the next Section, however, we review proposals that are specifically designed to observe and test GR effects using quantum systems based on atom interferometry.

%------------------------------------------------------------------------------------------------------------------------------------------------------------------%
\subsection{Testing General Relativity With Quantum Systems} \label{sec:TestingGR}
%------------------------------------------------------------------------------------------------------------------------------------------------------------------%

Atom interferometry is now starting to reach a precision such that it could, in principle, be used to test GR \cite{Jentsch2004,PhysRevD.73.022001,PhysRevLett.98.111102,PhysRevD.78.042003}. %These would be laboratory tests of GR, which could potentially allow for improved isolation of relativistic effects, in contrast to astronomical tests.  
Measurable effects of GR could include non-linear effects; velocity dependent forces; the equivalence principle; the Lense-Thirring effect; and GWs, the latter of which will be discussed in the following section.  In \cite{PhysRevD.73.022001,PhysRevLett.98.111102}, atom interferometry tests of GR were motivated by comparing certain GR and non-relativistic effects to determine if, in principle, the former could be  distinguishable from the latter.  Light-pulse atom interferometry was promoted to a relativistic setting with both the atoms and the light treated relativistically and all coordinate dependencies removed.  In particular, a semiclassical description was used where the free propagation of the atoms and the light was treated classically such that, in analogy with non-relativistic calculations, the propagation phase is proportional to the classical (but now GR) action, whereas the atom-light interaction is interpreted using non-relativistic QM but described in a covariant manner.  The atom interferometry was assumed to be placed in a Schwarzschild metric created by the Earth with a post-Newtonian expansion of the  metric utilized since the Earth's gravitational field is weak.  This post-Newtonian expansion was parameterized such that any effects beyond GR could be investigated.  Relativistic effects that were studied include non-linear terms, gravity from the atom's kinetic energy, and the falling of light.  The non-linear and velocity dependent effects were found to be measurable with an effective strength of $10^{-15}g$, and it was argued that such small accelerations could still be observable by implementing particular setups such that these relativistic effects can be isolated from the total phase shift.  For example, magnetic shielding, a rotation servo to null the Earth's rotation rate, strategically placed masses, and multiple atom interferometers in different configurations could be employed.  Furthermore, with advancements such as using entangled states and increasing the length of atom interferometers, it was argued that it would be possible to improve the precision by several orders of magnitude and exceed that of present astrophysical tests of GWs. In \cite{PhysRevLett.98.111102}, other metrics were also assumed, such as those created by a rotating planet and the expansion of the Universe, to determine if additional effects could be measurable (e.g.\ the Lense-Thirring effect, which was also discussed in \cite{Jentsch2004}).  

%Note that, although testing for effects beyond GR, the above quantum experiments won't be able to  distinguish GR from Newtonian gravity.  However, there have been recent proposals for interferometers that are specifically designed to allow for the distinguishability of the two theories by investigating their effects on quantum properties of a system.  
Using  atom interferometry to test GR would, in principle, demonstrate relativistic effects on quantum properties since they utilize quantum coherence.  However, atom interferometers could also be used to illustrate how GR affects quantum properties in a rather different setting: in \cite{Zych2011} an atom interferometer experiment was proposed where time dilation induced by different  gravitational potentials in the two arms of the experiment should, in theory, lead to a loss in visibility of the quantum interference pattern for the two paths. What makes this experiment unlike standard atom interferometer setups exploring gravity is that an intrinsic time-evolving degree of freedom of the matter-wave is used as a clock \cite{Zych2011,Sinha2011}, making time a `which path' witness.  There must, therefore, be a loss in the contrast of quantum interference due to Bohr's complementarity principle, leading to an interesting interplay between spatial and temporal degrees of freedom in the interference process. A proof-of-principle experiment was carried out for this type of proposal last year \cite{Margalit1205}. This experiment consisted of a $^{87}\mathrm{Rb}$ BEC  where the  $\mathrm{mF} = 1$ and $\mathrm{mF} = 2$ sublevels of the $\mathrm{F} = 2$ hyperfine state were used for the clock.  This clock ticks too slowly to observe time dilation and instead a vertical magnetic-field gradient is used to simulate a gravitational field by driving the clocks out of phase.  As anticipated, when the clocks are made to tick at different rates, there is a loss in the visibility of the interference pattern.  To test GR, however, certain hurdles will need to be overcome \cite{Arndt1168}.  Once this is achieved it would allow for experiments that are able to explore certain aspects of the interplay of GR and QM, such as the differing roles of time in the two theories, and if gravity has any role in the transference from the quantum to the classical world \cite{Pikovski2015}. 

An optical version of this type of experiment has also been proposed where a single photon is used rather than an atom, and the function of the `clock' is given by the photon's position along the interferometer's arm \cite{Zych2012}. Then, if the difference in the time dilations is comparable with the photon's coherence time, there would also be a loss in the visibility of the quantum interference brought about by the different paths. An advantage that this scheme has over using an atom interferometer is that even the measurement of the gravitationally induced phase shift would be of relevance since Newtonian gravity should not couple to a strictly massless system.

Note that all of these proposals use a framework of non-relativistic QM evolving via the Schr\"{o}dinger equation but with time, space, and any external gravitational fields displaying properties of GR.  For example, an atom would be described as a quantum system that evolves via the Schr\"{o}dinger equation (and is not considered to be an excitation of a quantum field) but with the temporal variable that enters into the Schr\"{o}dinger equation undergoing time dilation due to an external non-uniform gravitational field.  This fundamentally differs to the proposals that are outlined in Sections \ref{sec:corrections}-\ref{sec:RQTechnology} which use the framework of QFTCS and can consider extended rather than point-like systems.  %These proposals are also able to distinguish Newtonian gravity from GR but, in contrast to non-relativistic QM, QFTCS is a theory in which GR and the quantum world can both be incorporated at low energies. %relevant for current experiments.

Although tests that can distinguish GR from Newtonian gravity have not so far been carried out with atom interferometers, these devices have been used to test the Equivalence Principle  (EP) \cite{Fray2009,PhysRevA.88.043615,PhysRevLett.112.203002,PhysRevLett.115.013004}. These have so far been laboratory-based  investigations but space missions with significantly improved precision have also been proposed. For example, the STE-QUEST satellite mission, although not  not ultimately selected by the European Space Agency (ESA), was designed to test for the universality of free fall of matter waves by comparing the trajectory of two Bose-Einstein condensates of $^{85}\mathrm{Rb}$ and $^{87}\mathrm{Rb}$ \cite{Altschul2015501}; and  
the QTEST (Quantum Test of the Equivalence principle and Space Time) mission has been designed for the International Space Station (ISS) and will use two rubidium isotope gases with a  precision that is two orders of magnitude superior to the current limit on EP violations, and six orders of magnitude better than similar quantum experiments demonstrated in laboratories \cite{1367-2630-18-2-025018}.

%------------------------------------------------------------------------------------------------------------------------------------------------------------------%
\subsection{Gravitational Wave Detectors Using Quantum Systems} \label{sec:GWDetectors}
%------------------------------------------------------------------------------------------------------------------------------------------------------------------%

GWs were finally discovered by LIGO earlier this year  after decades of experimental work carried out by more than a thousand scientists and engineers \cite{PhysRevLett.116.061102}. 
% in a milestone moment for physics .  
The GWs were detected in the interference pattern created by a laser interferometer that used a power-recycled Michelson interferometer with Gires-Tournois etalon arms of 4\,km length.   The device is currently sensitive enough to be able to detect a change in the distance between the solar system and the nearest star to just the thickness of a human hair.  However, plans are currently under way to upgrade the device to further improve the precision.  This will be made possible by, for example, exploiting quantum properties such as squeezing the laser light \cite{AasiJ.2013,Oelker:14}.  In this case a quantum system would be used to probe purely relativistic effects with breathtaking  accuracy.

Apart from LIGO there are many other schemes designed to measure GWs.  Typically these either employ large optical interferometers similar to LIGO, or follow the pioneering experiments of Weber by utilising large metal (Weber) bars \cite{PhysRevLett.18.498}. In the past few years  there have also been proposals to use atom interferometers for the detection of GWs.% [MAYBE ADD AN EARLY REF ON THIS? - ISSUE IS MIGO].
For example, very recently the group of P. Bouyer has suggested using a chain of atom interferometer based gradiometers for the detection of GWs with frequencies between 0.3 - 3\,Hz \cite{PhysRevD.93.021101}. This particular regime has so far been inaccessible to earth-based GW detectors due to the influence of fluctuations in the terrestrial gravitational field, the so-called Newtonian noise (NN). The suggested experiment overcomes this noise by  utilising an array of atom interferometers in an appropriate configuration such that the noise can be rejected and the GW signal can be extracted by averaging over several realizations of the NN.  %Furthermore, since the array of atom interferometers are linked through %all interferometers  using 
%the same standing wave, which acts as a beam splitter, %The atom interferometers are based on three Bragg reflections acting as large momentum transfer beam splitters. 
%and the use of one common standing wave rejects fluctuations due to mechanical vibrations. 
%This setup would therefore act as a multiple differential gravimeter, which could also be used to determine a high order gravitational curvature [MIGA]. 
The separation chosen between the individual atom interferometers determines the Newtonian noise rejection efficiency. For a baseline length of 16.3\,km, 80 gradiometers, and separations of $\delta = 200$\,m, they predict an optimum sensitivity of $3\times 10^{-23} /\sqrt{\mathrm{Hz}}$ for a frequency of 2\,Hz.      

%The two principle sources of this NN are seismic waves propagating within the ground and air mass fluctuations in the nearby atmosphere. 

There have also been proposals to use atom interferometers in space for detecting GWs \cite{PhysRevLett.110.171102,2015arXiv150106797H}.  
%Hogan and Kasevich have proposed to use atom interferometers to improve GW detection in space based detectors .
For example, the proposal in \cite{2015arXiv150106797H} positions two light-pulse atom interferometers in separated spacecrafts with the relative acceleration measured using a Mach-Zehnder type configuration. The interferometers are operated with local lasers, which are phase-stabilized to a third laser beam linking the two spacecrafts. With separations of two spacecrafts on the order of $10^6$\,m, a passing GW would then affect the laser beam linking the two spacecrafts to a measurable level by creating a relative phase-shift between the two light pulse atom-interferometers.

In addition to atom interferometers, theoretical and experimental 
progress has also been made in developing superconducting
microwave cavities for the detection of GWs.  The Microwave Apparatus for Gravitational Waves Observation (MAGO) is based on the concept that a GW could induce an energy transfer between two levels of an electromagnetic resonator \cite{MAGO2005}, where the two levels are obtained in this case by coupling two identical high frequency spherical superconducting cavities.  The whole setup is only of the order of a metre long and works in the frequency range $10^3$-$10^4$\,Hz, offering a relatively cheap proposal to investigate the high frequency ($f \gtrsim 4 \times 10^3$\,Hz) GW spectrum.

%------------------------------------------------------------------------------------------------------------------------------------------------------------------%
\subsection{Massive Quantum Systems} \label{sec:Massive}
%------------------------------------------------------------------------------------------------------------------------------------------------------------------%

The previous sections reviewed quantum experiments that involve macroscopic distances such that relativistic effects from the Earth's gravitational field on quantum properties are becoming detectable.  For example, in Section \ref{sec:LongRange} we discussed proposals to distribute entanglement between Earth and satellite based stations and, in Section \ref{sec:QMClocks}, we reviewed how quantum clocks are reaching a precision such that relativistic effects could, in principle, be detected over macroscopic distances of the order of centimeters.  With further developments, these experiments will effectively be able to  probe low-energy consequences of quantum gravity where matter is quantized but the gravitational field in which they move is, to a very good approximation, described by GR.  This regime is expected to be well-described by QTFCS  and so, if observed effects are not in agreement with this theory's predictions, then this would challenge our current perceptions of gravity at the quantum scale.  So far we've only been able to test how Newtonian gravity affects quantum systems, and this has been in agreement with our expectations \cite{PhysRevLett.34.1472,Peters1999}.

This approach to investigating quantum gravity in the lab is in contrast to that where one would attempt to probe microscopic (Planck length) effects of gravity by, for example, sending elementary quantum particles to Planck scale energies.  Instead, rather than taking gravity to microscopic regimes where the quantum world dominates, the experiments reviewed in the previous sections are bringing quantum systems to our macroscopic world where gravity dominates.  In these experiments, quantum systems are brought to length scales where the effects from Earth's gravitational field are enhanced.\footnote{Even the quantum clock experiments of Section \ref{sec:QMClocks} are working at macroscopic scales but the length scales (cms) are much smaller than those of the experiments in Section \ref{sec:LongRange} due to their precision.}  However, another way to enhance gravitational effects is to increase the rest mass of the quantum system.  One could then, in principle, measure the gravitational field generated by a quantum system in the lab.  In particular, this would enable an experiment to measure the gravitational field that is generated by a quantum system in a spatial superposition.  From a quantum gravity point of view this should create a superposition of gravitational fields so that a nearby quantum system would be put into a superposition of motions \cite{DeWitt2011}.  Measuring this system would, therefore, inform us about a quantum theory of gravity. In particular, it could rule out theories in which gravity is predicted to be fundamentally classical and not quantum, since, in this case, the second quantum system would not be placed into a superposition of motions. 

%%%%%%%%%%%% LESS DETAIL %%%%%%%%%%%%%%%%%
%Unfortunately, performing this experiment is extremely challenging since, on the one hand, the gravitational field of small masses is difficult to detect and, on the other hand, the larger the mass of the system the more difficult it is to prepare in a spatial superposition.  However, there has been significant experimental progress on both sides of this problem in recent years, particularity with increasing the mass of systems in a superposition of states.

%%%%%%%%%%%%%% MORE DETAIL %%%%%%%%%%%%%
The notion that gravity is fundamentally classical, or emergent, has been growing in popularity in recent years with proponents pointing to the fact that the quantization procedure that has been successfully applied to the other forces cannot be straightforwardly applied to GR.  Furthermore, it is argued that, perhaps because gravity is not really a force according to GR but instead a manifestation of the curvature of spacetime, or perhaps because it is inherently non-linear in contrast to quantum theory and electromagnetism, no theory of quantum gravity exists in nature.  Such a fundamental view  of gravity where matter is quantized but gravity remains classical cannot be described by QFTCS since this theory neglects the gravitational field generated by a quantum system.  Instead the most natural option would be to extend QFTCS so that the gravitational field depends on the expectation value of the energy-momentum tensor of the quantum state.  That is, QFTCS would be extended with the semiclassical Einstein equations, which is a theory of semiclassical gravity that is usually considered to be a possible effective theory of quantum gravity at low-energies rather than a fundamental description of gravity.  Since the classical gravitational field depends on the wavefunction of a quantum system in this theory, by taking on a more fundamental nature, the theory would require the wavefunction to be interpreted as a real object and the classical gravitational field to be non-continuous \cite{DeWitt2011}.  It is also potentially inconsistent with observation within a certain interpretation of quantum theory \cite{PhysRevLett.47.979} and  plagued by superluminal signalling \cite{Pearle1996}, although this might be avoided if it can be combined with a mechanism of gravitationally induced collapse of the wavefunction \cite{Kibble1981,Pearle1996}.  

By measuring the gravitational field of a quantum system in a spatial superposition one should be able to rule out this semiclassical theory of gravity as a fundamental description of gravity interacting with the quantum world. Unfortunately, performing this experiment is extremely challenging since, on the one hand, the gravitational field of small masses is difficult to detect and, on the other hand, the larger the mass of the system the more difficult it is to prepare in a spatial superposition.  However, there has been significant experimental progress on both sides of this problem in recent years, particularity with increasing the mass of systems in a superposition of states. The current mass record for an object placed in a spatial quantum superposition over distances comparable to its size is 10,000\,AMU, which was demonstrated in macromolecule interferometry \cite{C3CP51500A,Cotter2015}. There has also been considerable interest in recent years in whether the use of levitated quantum nano-mechanical oscillators could significantly improve  this mass record  \cite{PhysRevLett.107.020405,PhysRevA.84.052121,Bateman2013,2016arXiv160301553P}.  Quantum objects with masses as large as $10^{11}$ and $10^{13}$ AMU have been prepared but so far the delocalizations achieved have been much smaller than their size \cite{OConnell2010,Teufel2011,Chan2011,RevModPhys.86.1391}.  However, proposals exist to significantly extend the delocalization of such systems.  For example, in \cite{2016arXiv160301553P} it is argued that, in principle, a micron-sized superconducting sphere of mass $10^{13}$ AMU could be prepared in a spatial superposition of the order of half a micrometer.  A principle aim of all these experiments investigating massive quantum systems is to determine if QM can be taken to macroscopic scales or whether it breaks down at some point due to, for example, a fundamental objective modification that results in a non-unitary evolution of the system and an effective collapse of the wavefunction.  In fact it has been suggested that such a process could be required for gravity to be reconciled with QM \cite{FeynmanLecturesGravity,Karolyhazy1966,PhysRevA.40.1165,Penrose1996}, with motivation coming from, for instance, the non-linearity of GR or the potential granularity of spacetime.  In particular, QFTCS with the semiclassical Einstein equations, and its non-relativistic counterpart the Schr\"{o}dinger-Newton equation, have inspired gravitationally induced collapse mechanisms \cite{DIOSI1984199,Penrose1996,Penrose1988,Penrose2014}. More quantum gravity-like scenarios have also been suggested, such as predictions that the wavefunction cannot be sustained in a superposition if the gravitational self-energy of the difference between the states is greater than the Planck energy \cite{Penrose1996}, or if gravitational fluctuations from granularity of spacetime or the cosmic micro-wave background may be responsible \cite{PhysRevLett.111.021302}. Furthermore, low-energy QM in the presence of the Earth's gravitational field, but with no additional extensions, has been considered, with an inherent decoherence process originating from time dilation \cite{Pikovski2015}.

It is possible that micro-mechanical devices could also help with the other side of the experimental challenge: measuring the gravitational field of very small masses \cite{2015arXiv150705733B,Jonas2016}.  In particular, in \cite{Jonas2016} a proof-of-principle experiment was outlined where a micro-mechanical device is resonantly driven by the gravitational field generated by a nearby oscillating source mass of just a few grams.  This source mass would be around three orders of magnitude smaller than what has been measured so far.  However, although this would represent substantial progress, there is still some way to go before the mass scale begins to close in on what can be put into a superposition.  Therefore, a top-down approach has been taken and the initial objective of this mechanical setup is to measure Newton's constant $G$.\footnote{A levitation version of this experiment using two cold superconducting spheres moving along parallel magnetic waveguides has also been considered in \cite{2016arXiv160301553P}}  This constant currently has the greatest uncertainty of all the fundamental constants with different groups obtaining values that are outside each other's error bars.  A revolutionary new way of measuring $G$ may, therefore, provide information on why the measurements differ so wildly and enable better characterization of the systematic errors in these measurements. 

Although the objective of this experimental approach is to measure effects that can be accounted for by, at first classical, and eventually quantum, Newtonian gravity, the incredible precision of these setups, and their expected rapid progress, potentially opens up the possibility to measure \textit{relativistic} effects of gravity.  So far GR has only been tested with astronomical source masses, such as the Earth and far-away stars and galaxies, whose motion we have no control over.  However, laboratory based experiments would provide completely new tests of GR since they would offer control over the motion and properties of the source mass.  By further using quantum systems as test masses, one might  then be able to measure relativistic effects on certain quantum properties in a highly controlled setting.

%------------------------------------------------------------------------------------------------------------------------------------------------------------------%
\subsection{Quantum Experiments Probing the Dynamical Casimir Effect} \label{sec:DCE}
%------------------------------------------------------------------------------------------------------------------------------------------------------------------%

If the boundary conditions, dispersion law, or background of a quantum field are quickly varied in time such that the system is unable to adiabatically follow the instantaneous ground state, then correlated pairs of excitations are created out of the vacuum state.  This is the DCE \cite{MooreDCE,Fulling393}, which is a clear relativistic effect and is typically described in terms of the production of photons from the electromagnetic vacuum between two mirrors when one of the mirrors undergoes a rapid acceleration. This effect has received numerous theoretical studies since its incarnation but it has only recently  received experimental tests.%verification.  
 This is principally because, in its typical formulation, the rate of photon production is only non-negligible when the acceleration is extremely large, which makes the use of massive mirrors particularly challenging \cite{PhysRevLett.96.200402,refId0}. However, the DCE is not just restricted to this setup of moving mirrors, and this has led to a number of alternative experimental proposals.

In 2011, the DCE was observed for the first time by   modulating the electrical properties of a cavity \cite{Wilson2011}.  The system consisted of a coplanar waveguide (CPW) terminated by a superconducting quantum interference device (SQUID), and the boundary conditions were rapidly changed by  modulating the inductance of the SQUID.  The changing inductance can be described as a change in the electrical length of the transmission line and provides the same time-dependent boundary condition as the idealized moving mirror \cite{Fulling393,PhysRevLett.77.615}. The effective boundary can be changed rapidly enough in this case such that photon production from the vacuum is experimentally detectable.  %Furthermore, in addition to observing the creation of real photons, the group also detected two-mode squeezing in the emitted radiation, which is a smoking gun signature of the quantumness of the radiation.

In the same year, the DCE was also demonstrated using a Josephson metamaterial consisting of an array of 250 SQUIDs embedded in a microwave cavity \cite{Lahteenmaki12032013,2014arXiv1401.7875L}.  In this setup  the background in which the field propagates was quickly changed in time by periodically changing the index of refraction, resulting in the generation of quantum-correlated photons from the vacuum.   
%An interesting feature of this scheme compared to the above alternative realization that changes the boundary conditions in a transmission line,  is that there is a resonant enhancement of the DCE due to the cavity \cite{Lähteenmäki12032013}.  
%Both of these groups have thus theoretically and experimentally proven the relation between a single SQUID and a relativistically moving mirror.  The theoretical mapping between two SQUIDs and a relativistically moving cavity was then later shown in \cite{PhysRevLett.110.113602}. The experimental platform implemented in this case consists of arrays of hundreds of SQUIDs, and the refractive index for the microwave photons can be controlled externally by bias magnetic fields applied to the SQUID loops. Therefore, with appropriate sample design, the effective spacetime metric of the photonic excitations can be modulated either locally (below the wavelength size) or globally. Besides the demonstration of the dynamical Casimir effect, the work presented in \cite{PhysRevLett.110.113602} also provided preliminary results on the local modulation of the speed of light, showing conclusively that the local curvature of the effective spacetime has measurable effects. 
This experiment and that of \cite{Wilson2011} have demonstrated the potential power of utilising relativistic effects in quantum information tasks, which is discussed further in Section \ref{sec:RQTechnology}.
%For example, to create two-mode squeezing.
However, in order to use these effects for quantum computing, more complex states need to be generated \cite{2015arXiv151205561L}. Recently, a step towards this has been performed by generating coherence as well as squeezing correlations in tripartite states by double parametric pumping of a superconducting microwave cavity in the ground state \cite{2015arXiv151205561L}.   Furthermore, in \cite{Bruschi2016} continuous variable Gaussian cluster states where considered to be generated by  simulating relativistic motion in superconducting circuits.  Proposals for simulating the DCE in other systems have also considered. In particular, an acoustic analogue of the DCE was investigated in a BEC  but the temperature was too high to observe quantum effects \cite{PhysRevLett.109.220401}.  

%In particular, two-mode squeezing \cite{PhysRevLett.111.090504}, beam-splitting   and multipartite quantum gates \cite{PhysRevD.86.105003} were theorised in addition to a process in which continuous variable Gaussian cluster states where generated, which are universal resources for universal one-way quantum computation.  All of these processes can be  implemented experimentally using superconducting circuits

As well as paving the way for quantum information tasks based on the DCE and similar processes, these experiments have also demonstrated the potential of SQUIDs and BECs for simulating effects of curved spacetime on quantum properties  since the equivalence principle links accelerations and gravity.  Proposals for experiments that simulate QFTCS effects in this way are discussed further in Section \ref{sec:corrections}.

%------------------------------------------------------------------------------------------------------------------------------------------------------------------%
\section{Quantum Field Theory in Curved Spacetime} \label{sec:QFTCS}
%------------------------------------------------------------------------------------------------------------------------------------------------------------------%

In this section we briefly review QFTCS and introduce the motivating factors behind using this framework to analyse experiments that are approaching relativistic regimes, such as those  reviewed in Section \ref{sec:QMExpsRelativiticRegimes}.  

\subsection{Background}

Quantum Field Theory (QFT) is currently our best attempt at merging special relativity and quantum physics into a coherent framework, and QFTCS represents the first step in extending this highly successful theory to curved spacetime  \cite{Birrell:Davies:1984}. In QFTCS, matter obeys QFT but the spacetime in which the quantum field of matter propagates is classical and is not affected by it.  That is, spacetime is taken to obey the rules of GR but with its curvature pre-assumed and not influenced by quantum matter, whereas matter obeys QFT but is influenced by the curvature of the classical spacetime.  This theory is expected to offer an effective description of the interplay of quantum physics and GR at low-energies i.e.\ below the Planck scale, and is therefore a limiting case of quantum gravity.  Predictions of this theory, such as the Hawking effect \cite{Hawking:1974,Hawking1975} and particle creation from an expanding Universe \cite{PhysRev.183.1057,PhysRevD.3.346}, are among the most dazzling and challenging of modern physics, and in the past few decades an overwhelming amount of research has been dedicated to understanding how to demonstrate these predictions.  This has principally included astrophysical and cosmological studies but there has recently been a growing interest in complementing these studies with laboratory based setups.  As yet there have been no direct  measurements of QFTCS but quantum experiments operating at relativistic gravity regimes offer hope that such tests could be realized in the near future.

\subsection{Motivation}

The quantum experiments that were reviewed in  Section \ref{sec:QMExpsRelativiticRegimes}, with the exception of those in Section \ref{sec:DCE}, are all based on non-relativistic QM.  That is, their dynamics are described by the Schr\"{o}dinger equation, which is inherently non-relativistic.  However, as these experiments approach relativistic regimes, it will be appropriate to describe these systems in terms of QFT and, as relativistic gravity becomes relevant, in terms of QFTCS. Experimental signatures of QFTCS will then enable us to learn about the effects of gravity and motion on quantum resources, such as entanglement, at low-energies.  From this we can  also formulate predictions about how the performance and precision of future technologies will be modified by relativistic effects.  Such effects will need to be corrected for or new methods must be found to minimize them.  Recently there have been a number of theoretical studies that have analysed how the notion of relativistic gravity can affect the performance of certain quantum technologies.   In particular, corrections due to relativistic effects have been analysed using QFTCS for quantum technological systems that involve long-range quantum communication protocols, quantum teleportation, and quantum clocks.  Section \ref{sec:corrections}  reviews these recent theoretical studies.  %Additionally, it also discusses proposals that have been purposely designed to measure certain predications of QFTCS using current technology.

\subsection{QFTCS in Experiments and Technology}

Observing influences of QFTCS effects on quantum technologies would provide the first measurements of QFTCS and would thus give credence to effects such as Hawking radiation.  As well as providing corrections to existing non-relativistic protocol designs, our improved understanding of these effects could also open up the possibility for developing new technologies that are not only  compatible with the notion of spacetime but also directly exploit relativistic effects to improve their capabilities.  Examples of such \textit{relativistic} quantum technologies have  been developed by applying  Quantum Information Theory (QIT) techniques to QFTCS.  These proposed technological devices include an accelerometer; a device that can measure gravitational properties of the Earth; and a GW detector.  %Interestingly, theoretical work has demonstrated that such spacetime probes can offer a performance that is superior to their non-relativistic counterparts.  
%Realisation of these devices would represent the first time that the theory of relativity  has been used in the fundamental function of a technology, thus opening up a new era in technology. 
Section \ref{sec:RQTechnology} is dedicated to reviewing these proposed relativistic quantum devices. 

%As well as improving upon the performance of non-relativistic quantum technologies in industrial applications, 
These new spacetime probes could also be used to explore the overlap of quantum physics and GR, potentially providing experimental insights into quantum gravity.  A significant advantage that these experiments offer over astrophysical and cosmology searches for signatures of quantum gravity is that they would be based in a laboratory, providing a controlled setting and repeatability.  Observations of QFTCS effects in these laboratory-based experiments would also offer direct tests of QFTCS in contrast to any astrophysical or cosmological studies as well as analogue gravity experiments where only analogue effects can be seen \cite{PhysRevLett.46.1351}.  On the other hand, any  violation of QFTCS would offer a tantalising signature of the need for a radically new approach to quantum gravity.

%------------------------------------------------------------------------------------------------------------------------------------------------------------------%
\section{Relativistic Effects of Gravity on Quantum Properties}  \label{sec:corrections}
%\section{Relativistic gravity affects quantum properties}  \label{sec:corrections}
%------------------------------------------------------------------------------------------------------------------------------------------------------------------%

In this section we review recent studies that have utilized a QFTCS framework to show that the performance of a range of existing quantum technologies will be modified by relativistic effects of gravity.  Such quantum technologies include long-distance quantum communication protocols that use and don't use  entanglement, as well as quantum clocks.  These are all reviewed in Sections \ref{sec:GAffectsEntanglement}-\ref{sec:GAffectsQMClocks}.

%------------------------------------------------------------------------------------------------------------------------------------------------------------------%
\subsection{Gravity and Motion Affect Entanglement} \label{sec:GAffectsEntanglement}
%------------------------------------------------------------------------------------------------------------------------------------------------------------------%

In \cite{PhysRevD.85.061701}, a quantum information %gedanken
experiment was theorized that involves the entanglement of two scalar quantum fields. % each consisting of two quantum information states.  
Each  scalar field is placed in a cavity and the full system is initially prepared in the  maximally entangled Bell state $|\psi\rangle =  (|0\rangle |0\rangle + |1\rangle_k |1\rangle_k) / \sqrt{2}$ where $|0\rangle$ denotes the vacuum and $|1\rangle_k$ denotes  the one-particle state of mode $k$.  After the experiment is prepared, one of the cavities undergoes a non-uniform acceleration for a certain period of time, and this period of non-uniform acceleration was found to cause the entanglement between the cavities to  degrade.  The origin of this loss in entanglement is in fact  the same physical mechanism as the particle generation in the dynamical Casimir and Unruh-Hawking radiation effects.  That is, it is a purely QFTCS effect.

\begin{figure}[t!]
	\centering
	\includegraphics[width=0.3\textwidth]{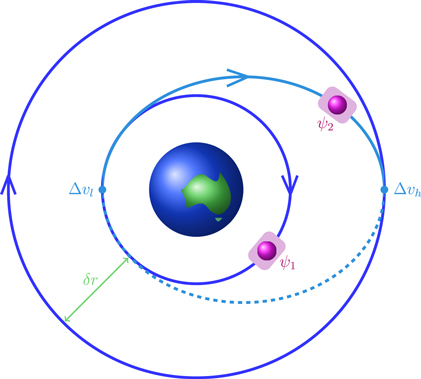}
	\label{DishwahserTabletsLabel}
	\vspace{0.25cm}
	\caption{\small Experiment proposed in \cite{Bruschi:Sabin:2014}.  Two BECs inside separate satellites are entangled while both are in the same circular LEO orbit. One of them then undergoes an acceleration for a finite time in order to change to a different circular orbit, which translates in a change of gravitational field felt by the satellite during this process.  The entanglement between the BECs is found to be degraded to an observable level.}
\end{figure}

This theoretical experiment illustrates how quantum information protocols that involve entanglement, such as quantum teleportation, can be affected by non-uniform accelerations\footnote{For uniform accelerations see e.g.\ \cite{PhysRevLett.91.180404,Adesso2012}} and, in principle, from the equivalence principle, by changing gravitational fields. Scenarios of this type are, therefore, relevant to the types of experiments reviewed in Section \ref{sec:LongRange} where the quantum communication and teleportation protocols take place over large distances where the gravitational field of the Earth is non-constant.  It is also relevant to the type of experiments discussed in Section \ref{sec:GMetrology} that are  proposed to measure gravitational effects  using entangled systems. Although presently unobservable in these quantum technology experiments,  as they become more precise, and as the distances involved increase, this degradation in entanglement will become more relevant to their operation.  However, these quantum experiments are not designed to look for such effects.  Instead they are devised to advance quantum technology. Therefore, it is possible that, if experiments are purposely designed such that these QFTCS effects are enhanced, degradation in entanglement could be observed for non-uniform accelerations or gravitational fields using current technology.  Such an experiment was proposed in \cite{Bruschi:Sabin:2014} where it was shown that, by entangling the phononic modes of two BECs that are placed in two separate satellites, the entanglement would be degraded to an observable  degree if one of the satellites subsequently moved to a new orbit. An illustration of this experiment is presented in Figure \ref{DishwahserTabletsLabel}.  The reason that this effect is predicted to be currently observable using BECs and not, for example, with a cavity of light, is that the speed of propagation of phonons of BECs is considerably smaller than that of light \cite{Bruschi:Sabin:2014}.% and the BEC in some sense thus acts a dielectric.

The effects of relativistic motion on quantum teleportation can also be simulated in the laboratory  using experiments based on superconducting resonators with tunable boundary conditions \cite{PhysRevLett.110.113602}.  The superconducting experiments reviewed in Section \ref{sec:DCE} proved the relation between a single SQUID and a relativistically moving mirror.  To simulate a relativistically moving cavity, one instead has to use two SQUIDS \cite{PhysRevLett.110.113602}.  The experimental platform implemented in this case consists of arrays of hundreds of SQUIDs, and the refractive index for the microwave photons can be controlled externally by bias magnetic fields applied to the SQUID loops. Therefore, with appropriate sample design, the effective spacetime metric of the photonic excitations can be modulated either locally (below the wavelength size) or globally.   By simulating accelerations of a quantum field cavity, and thus gravitational fields, it was shown that the relativistic effects on a quantum teleportation protocol would be sizeable for realistic experimental parameters.

%------------------------------------------------------------------------------------------------------------------------------------------------------------------%
\subsection{Gravity and Motion Affect Quantum Communication Protocols} \label{sec:GAffectsCommunication}
%------------------------------------------------------------------------------------------------------------------------------------------------------------------%

\begin{figure}[t!]
	\begin{center}
		%\subfigure[]{
		\resizebox*{7cm}{!}{\includegraphics{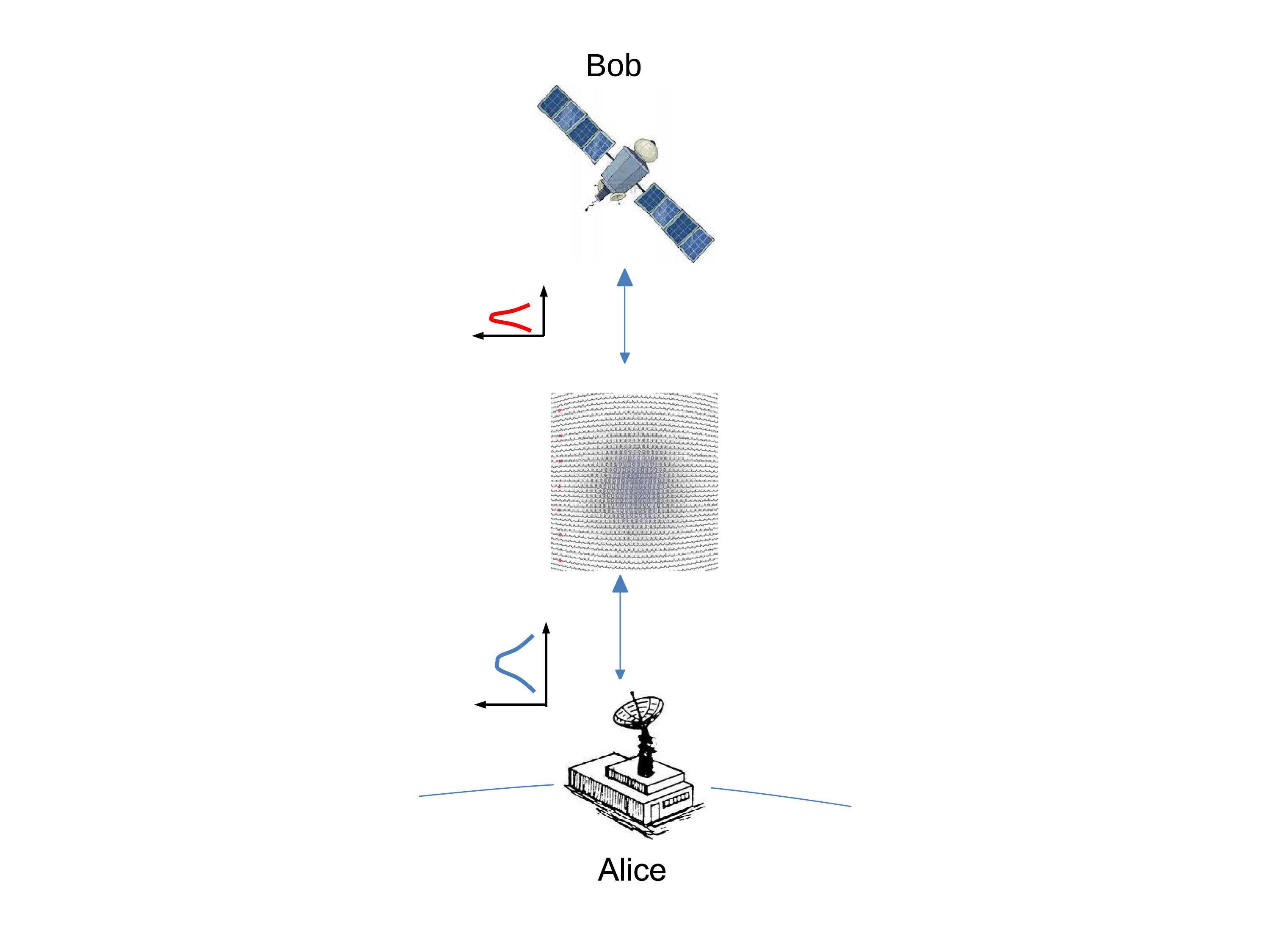}}\hspace{5pt}
		%\subfigure[]{\resizebox*{7cm}{!}{\includegraphics{imagekerr2}}\label{RotatingPlanet.b}}
		\caption{\small A quantum communication protocol performed between the Earth and a satellite.  Here Alice, located on Earth, prepares and sends photons to Bob who is located on a satellite.  The photon that is created
			by Alice has certain characteristics, such as peak frequency and
			bandwidth, that change once the photon is received by Bob since he at a different gravitational potential to Alice. Figure taken from \cite{Bruschi:Ralph:2014}.
		}
		\label{Communication}
	\end{center}
\end{figure}

Experiments and proposals were reviewed in Section \ref{sec:LongRange} that send pulses of light from the Earth to a satellite in a quantum communication protocol that could one day be used for a global quantum communications network. Treating this protocol within a QFTCS framework, it has been shown that the photons, which are never monochromatic in practice and so are modelled as peaked wavepackets, are red-shifted and thus lose energy as they travel through the Earth's non-uniform gravitational field \cite{Bruschi:Ralph:2014}. The satellite will, therefore, receive a wavepacket with a different frequency distribution to the one prepared and sent from the Earth.  That is, as illustrated in Figure \ref{Communication}, the quantum communication scheme will differ from one performed in flat spacetime. 
In \cite{PhysRevD.93.065008}, these relativistic effects were also taken into account in a scheme for near-Earth orbit satellite-based quantum clock synchronization.%, with   atmospheric dispersion cancellation by taking into account the spacetime background of the Earth.

%As these types of experiments become more precise and involve larger distances, this  discrepancy between a protocol performed in curved spacetime compared to flat spacetime could, in principle, be observable.  Not only would this form a test of QFTCS, it might also require that future quantum communication protocols need to take into account such effects or apply techniques in order ro compensate for the effects, such as potentially utilising a local oscillator \cite{}.  How these effects would impact such communication protocols, such as an entanglement distribution protocol and a continuous variable QKD protocol, was studied in \cite{Bruschi:Ralph:2014}.  In particular, it was shown that an Earth-to-space QKD system that relies on entanglement distribution using photons could have an additional contribution to its quantum bit error rate (QBER) as high as 0.7\% as a consequence of space-time curvature, which should be observable with current technologies \cite{Bruschi:Ralph:2014}.

%------------------------------------------------------------------------------------------------------------------------------------------------------------------%
\subsection{Gravity and Motion Affect the Precision of Quantum Clocks} \label{sec:GAffectsQMClocks}
%------------------------------------------------------------------------------------------------------------------------------------------------------------------%

As reviewed in Section \ref{sec:QMClocks},  clocks that are based on the optical transitions of ions or neutral atoms in optical lattices are achieving unprecedented levels of precision and accuracy. Current designs of quantum clocks are formulated within the framework of non-relativistic QM, and the measured relativistic effects such as time dilation can be explained by treating the clocks as point-like classical systems.  However, according to QFTCS, as clocks that exploit quantum properties, such as entanglement, become more precise then relativistic effects on quantum properties will need to be taken into account, and such modifications cannot, in practice, be accounted for by purely classical or non-relativistic quantum frameworks. %As argued in Section \ref{sec:}, the most consistent theory we have for analysing such effects at the energies at which the experiments operate is QFTCS.
This was illustrated in \cite{Lindkvist2015}, which considered a quantum field description of a quantum clock.  In this case a single mode of a scalar quantum field constrained to a cavity was used to model a clock, with the phase of the mode used as the pointer of the clock.\footnote{A more rigorous description of such a QFT clock is expected soon where the measurement readout of time will be taken into account.} This fundamental model is both necessarily quantum and relativistic.  That is, all quantities are compatible with Lorentz invariance and the notions of probability.  Furthermore, its motion through spacetime can be properly described.    For example, for uniform acceleration it was found that the rate of the cavity clock would be modified. This is due to the fact that, unlike previous studies which assumed the clock to be point-like, in this approach the clock has a length and different points in the cavity experience different proper times.  This effect can be understood classically. However, for changes in acceleration there is an additional phase shift due to mode-mixing and particle creation, which affects the precision of the clock and is a purely QFTCS effect.  Using the equivalence principle it was argued that the precision of the clock should also be affected by non-uniform gravitational fields.  Since in realistic settings a gravitational field is never stationary due to local effects such as seismic activity, this loss in precision would have to be accounted for in an ultra-precise quantum clock.  

In \cite{Lindkvist2015} it was shown that these results could, in principle, be simulated in the laboratory using superconducting quantum circuits.  Furthermore, by implementing two clocks with superconducting circuits, one would also be able to simulate the twin paradox scenario with velocities as large as a few percent of the speed of light \cite{PhysRevA.90.052113}.  Such velocities would allow one to investigate QFTCS effects, and theoretical results indicate that the time dilation in the paradox is altered due to quantum particle creation and for different cavity lengths \cite{PhysRevA.90.052113}.

%------------------------------------------------------------------------------------------------------------------------------------------------------------------%
\section{Using Quantum Properties to Measure Relativistic Effects \\ of Gravity} \label{sec:RQTechnology}
%\section{Using quantum properties to measure relativistic gravity} \label{sec:RQTechnology}
%------------------------------------------------------------------------------------------------------------------------------------------------------------------%

The previous section considered how relativistic effects of gravity can modify current quantum technologies such that their performance is degraded, requiring corrections to be made to their predicted outcomes.  In this section we instead look at the potential benefits of the relativistic effects.  An immediate possibility is to exploit the observations of relativistic effects on quantum properties to measure certain aspects of spacetime.  This notion has culminated in the creation of a new research field, relativistic quantum metrology, which investigates how devices based on QFTCS can be used to study the relativistic nature of spacetime and how these new quantum devices can enhance measurements of certain properties of spacetime.

%------------------------------------------------------------------------------------------------------------------------------------------------------------------%
\subsection{Relativistic Quantum Metrology} \label{sec:RQMetrology}
%------------------------------------------------------------------------------------------------------------------------------------------------------------------%

In quantum metrology, quantum properties are exploited to achieve greater statistical precision than purely classical strategies when determining a parameter of the system that is not an observable \cite{Braunstein:Caves:1994}.  Examples of such parameters include temperature, time, acceleration, and coupling strengths.  As with any measurement strategy, we can divide the process into the following phases: input of the initial state, evolution through a channel, and measurement of the output state.  The parameter to be estimated typically controls the evolution of the input state and is encoded in the output state. One then looks for the most optimal measurement strategy achievable such that the error in the estimation process is minimized.  

In classical physics, the error of a measurement can, in general, scale at most as $1/\sqrt{M N}$ on average, where $N$ is the number of input probes used in each estimation process, and $M$ is the number of times this estimation process is repeated.  %Therefore, $n:=NM$ is the total number of input probes used in this classical experiment.  
This scaling is just a consequence of the central limit theorem.  However, it has been shown that the `quantumness' of physical states and resources, such  as entanglement, can be exploited in order to achieve what is known as the quantum or Heisenberg scaling $1/ (\sqrt{M} N)$ where $N$ is, for example, the number of entangled input probes used in each estimation process, and $M$ is again the number of times this process is repeated \cite{PhysRevLett.77.2352}.  There is, therefore, an enhancement by a factor of $\sqrt{N}$ in the quantum case, and the overall scaling is a consequence of the Heisenberg uncertainty relation. This improved scaling can become extremely advantageous when the available number of input probes is high, and provides a signature of genuine quantum effects.  

Quantum metrology is being investigated in advanced technological proposals aimed at ultra-precise measurements of gravity such as the gravitational potential and the detection of GWs \cite{Jaekel1990,Schnabel2010,Oelker:14}. These systems have been designed using non-relativistic QM and thus utilize non-relativistic quantum properties.  However, recently quantum metrology has been applied to the field of QFTCS  to create a new line of research `relativistic quantum metrology' that investigates measuring spacetime quantities such as relative distances, proper times and gravitational field strengths by exploiting relativistic effects on quantum systems \cite{PhysRevLett.105.151301, 2011arXiv1108.5220D,PhysRevA.88.052112,Ahmadi:Bruschi:2014:v2,PhysRevA.89.042336}.  Excitingly, this has lead to the development of novel `relativistic' quantum devices that, in certain cases, appear to improve upon the precision of their non-relativistic counterparts \cite{Ahmadi:Bruschi:2014:v2,Sabin2014}.  Proposed relativistic quantum devices include an accelerometer \cite{Ahmadi:Bruschi:2014:v2}; a GW detector \cite{Sabin2014}; and a device that can measure Schwarzschild properties of the Earth \cite{Bruschi:Datta:2014,Kohlrus2015}, which are all reviewed in the following subsections.  Other investigations include how two entangled atoms could be used to detect spacetime curvature through the Casimir-Polder interaction \cite{2016arXiv160507350T},  how entanglement can enhance the precision of the detection of the Unruh effect using accelerated probes \cite{Wang2014,Tian2015,2016arXiv160301122T}, and how the expansion rate of the universe in the Robertson-Walker metric  for Dirac fields can be estimated \cite{Wang2015390}.  

%------------------------------------------------------------------------------------------------------------------------------------------------------------------%
\subsubsection{Measuring Schwarzschild Properties of the Earth} \label{sec:Schwarzschild}
%------------------------------------------------------------------------------------------------------------------------------------------------------------------%

Using relativistic quantum metrology techniques, it has been demonstrated that  the effects of gravity on propagating wavepackets can be utilized to measure the Schwarzschild radius of the Earth, the angular momentum of the Earth, or the relative distance between the Earth and a satellite \cite{Bruschi:Datta:2014,Kohlrus2015}.  In this scheme a wavepacket of light is sent from the Earth to a satellite and the parameters of interest are estimated from comparing the wavepacket's final state to the state of a wavepacket that has not undergone the trajectory.  This is made possible because, due to QFTCS effects, the background curvature becomes imprinted in the propagating photon or light pulse. The effects of spacetime on the photon can then be treated as a channel and one is able to employ a relativistic quantum metrology scheme to estimate the desired parameters. A lower bound is  obtained on the error of the estimation that depends on the spacetime parameters, the number of times the processes is repeated, and on the  squeezing that can be chosen for the initial state.  Given the amount of interest in space-based quantum protocols, and the rate of development of the technologies \cite{Dequal:Vallone:2016}, it is conceivable that experiments of this kind will be feasible within the next decade.

%------------------------------------------------------------------------------------------------------------------------------------------------------------------%
\subsubsection{An Accelerometer} \label{sec:Accelerometer}
%------------------------------------------------------------------------------------------------------------------------------------------------------------------%

In a BEC, phonons propagate on an \textit{effective} metric created by the condensate \cite{Ahmadi:Bruschi:2014:v2}.  However, as well as depending on BEC parameters such as condensate density and velocity flows, this effective metric also depends on the real spacetime metric \cite{Fagnocchi2010,Visser2010}.  In fact it depends on the real metric in a conformal way such that, if the velocity flows are neglected, the phonons behave  like unpolarized photons propagating in a relativistic field but at a highly reduced velocity.  In \cite{Ahmadi:Bruschi:2014:v2} this was utilized to demonstrate that, by oscillating the BEC, phonons will be created in a dynamical Casimir-like process.  However, unlike for a electromagnetic cavity, the frequencies of oscillations required to create measurable effects are very small, being that of typical phonon frequencies i.e.\ 10-1000\,Hz.  By creating a squeezed state of the phonons, allowing the phonons to freely propagate for a period of time, and then measuring the final state, one is  able to apply relativistic metrology techniques to obtain an estimate on the non-uniform motion of the cavity.
This, therefore, enables the design of a BEC accelerometer, and plans are currently under way to also design a BEC gravimeter using similar principles. Intriguingly, by comparing the accuracy of this relativistic device to the state-of-the-art, an improvement by two orders of magnitude is observed , demonstrating the potential in exploiting relativistic effects on quantum properties.

\subsubsection{A Gravitational Wave Detector} \label{sec:RGWDetector}
%------------------------------------------------------------------------------------------------------------------------------------------------------------------%

The proposal for the accelerometer reviewed in the previous section used the  fact that real spacetime distortions can create phonons of a BEC.  From the equivalence principle, it should, in principle, also be possible to use this in the design of a quantum gravimeter and  a GW detector.  The latter was illustrated in \cite{Sabin2014}  where it was shown that GWs will resonate with modes of the phononic field.  Since phonons can have frequencies of order 10-1000\,$\mathrm{Hz}$ in BEC experiments, the GWs that could, in principle, be detected would be those with a similar frequency range.  Such GWs are predicted to be generated by the merging of binary neutron stars and black holes systems, as well as potentially from continuous sources such as rotating neutron stars or pulsars.  The frequency range of 10-1000\,$\mathrm{Hz}$ overlaps with that of LIGO, and initial calculations using relativistic quantum metrology suggest that the strain sensitivity could be at least, in principle, comparable to aLIGO.  The detector proposed in \cite{Sabin2014} would also be considerably smaller, cheaper, and simpler to build than current laser interferometers that can detect GWs.  Among other benefits, this would enable many experimental groups to employ GW detectors, allowing for better determination of true GW signals through near-simultaneous measurements and precise positioning of the source of the signals.

Note that the GW detectors described in Section \ref{sec:GWDetectors} that are also based on BECs are designed using a fundamentally different concept to the GW detector described here, and work in a completely separate frequency regime.  These are atom interferometers and make use of non-relativistic QM and phase estimation techniques applied to  the condensate atoms such that one can, in principle, detect GWs with frequencies between $10^{-5}$ to $10^{-2}$\,Hz  with strain sensitivities of the order of $10^{-18}$\,$\mathrm{Hz}^{-1/2}$. In contrast, the GW detector described here uses state discrimination techniques applied to the phononic excitations of the BEC in a QFTCS description, which, in principle, should allow for strain sensitivities of the order of $10^{-27}$\,$\mathrm{Hz}^{-1/2}$ \cite{Sabin2014}.  Furthermore, whereas the frequency range of the GWs detected by the atom interferometers correspond to waves generated by small mass binary systems, the GW detector described here could, theoretically, detect GWs in the kHz regime, which  would  deepen our understanding of neutron stars by gathering the information necessary to describe their equation of state. This information would further our understanding of the  origins, evolution and extent of our universe since it would enable cosmologists to  compute distances that are key in the study of the cosmological constant and of dark matter.

\section{Conclusion} \label{sec:Summary}

Quantum experiments have undergone extraordinary developments in recent years. %In particular, the advancement in their precision has been remarkable with, for example, the recent demonstration of time dilation over just 33\,cm \cite{} using a quantum clock.  
These experiments are constantly probing and pushing science into new and unknown territories, such as investigating entanglement over terrestrial distances and superpositions over macroscopically sized objects. Excitingly, the constant progress provided by the successful accomplishments of these experiments implies that they are rapidly approaching a regime in fundamental physics that has yet to be explored.  This is the regime in which general relativistic effects of gravity will begin to influence quantum properties such as entanglement. The fact that quantum experiments are starting to approach such a regime is most clearly illustrated by the experiments that are attempting to prepare an entangled state between an Earth based station and a satellite for proof-of-principle experiments of a global quantum communications network.  Here distances are being used where GR already has to be taken into account in classical technology such as GPS.

In this article we have reviewed the current types of quantum experiments that are beginning to approach this relativistic regime, and have emphasized that the most appropriate theory for describing such experiments will be QFTCS.  As reviewed in Section \ref{sec:corrections}, recent work that has applied QFTCS to QIT has shown that relativistic effects of gravity can have a detrimental impact on the performance of such quantum devices, requiring corrections to be made in the predicted outcomes of quantum experiments.  However, as discussed in Section \ref{sec:RQTechnology}, rather than just having a negative impact, these relativistic effects can also be exploited to design quantum devices that can directly measure spacetime effects.  One such promising quantum device is a GW detector that is built from a BEC and operates in the high frequency regime where one expects to see binary black hole and neutron star mergers as well as, in principle, rotating neutron stars and pulsars.  Amazingly, this would mean that relativistic effects could be seen in an object the size of just micrometers.  As reviewed in Section  \ref{sec:RQMetrology}, this GW detector is just one example of how QFTCS can be utilized to develop new microscopic quantum devices capable of probing spacetime.  Technology of this kind could, in principle, provide deeper insights into our Universe through gravitational interaction.

%Interestingly, the strain sensitivity of this device is expected to be at least comparable to LIGO but advantages would include the fact that it would be smaller, cheaper and simpler to build, enabling  many experimental groups to implement such detectors.  This would then allow for pinpoint accuracy in the positioning of the sources, enabling a high performance GW telescope.

In summary, we expect to soon be embarking upon a new revolution in physics where we will observe the interplay of the quantum world and GR for the first time.  Rather than investigating GR at small distances, which would require huge advancements in technology, this interplay would involve classical GR affecting quantum properties in a laboratory setting.  The effects must be predictions of a limiting case of quantum gravity, and QFTCS is currently our best attempt at this.  Observing predictions of QFTCS could, therefore, provide guidance for theories of quantum gravity and the microscopic structure of spacetime.  %In particular, it would provide credence for effects such as Hawking radiation and its thermodynamic connections.

\section*{Acknowledgements}

We would like to acknowledge Advances in Physics:\hspace{1mm}X, for which an invited review was written that forms a major part of this manuscript.  D.E.B. also wishes to acknowledge hospitality from the University of Vienna.

\section*{Funding}

This project has been supported by EPSRC grants EP/M003019/1 and EP/K023624/1, and by the European Commission grant QuILMI - Quantum Integrated Light Matter Interface (No. 295293).  D.E.B., I.F. and R.H. would also like to acknowledge the John Templeton Foundation grant `Leaps in cosmology: gravitational wave detection with quantum systems' (No. 58745).

%\bibliographystyle{tADP}
%\bibliography{GravityintheLabReview}

%\sloppy %fix up any links that spread accross margins, but has poor spacing at times

\printbibliography
%\bibliography

\end{document}